\input harvmac
\input tables

\Title{hep-th/yymmnn SCIPP 96/26}
{\vbox{\centerline{Phenomenology of Strongly
Coupled}
\centerline{ Heterotic String Theory}}}
\bigskip
\centerline{Tom Banks }
\smallskip
\centerline{\it Department of Physics and Astronomy}
\centerline{\it Rutgers University, Piscataway, NJ 08855-0849}
\smallskip
\centerline{Michael Dine}
\smallskip
\centerline{\it Santa Cruz Institute for Particle Physics}
\centerline{\it University of California, Santa Cruz, CA   95064}
\bigskip
\baselineskip 18pt
\noindent

This is the text of a talk given at the Inaugural Conference of the Asia
Pacific Center for Theoretical Physics, Seoul, Korea, June 9, 1996.
If nature is described by string theory, and if the
compactification radius is large (as suggested by
the unification of couplings), then the theory is in a regime
best described by the low energy limit of $M$-theory.
We discuss some phenomenological aspects
of this view. The scale at which conventional quantum field theory
breaks down is of order the unification scale and consequently 
(approximate)
discrete symmetries are essential to prevent proton decay. There
are one or more light axions, one of which solves the strong CP problem.
Modular cosmology is still problematic but much more complex than
in perturbative string vacua.

\Date{6/96}

\def\Meleven{M_{11}}

\def\Releven{R_{11}}
\newsec{Introduction}

Before the advent of string duality, weakly coupled heterotic string
theory was the best string theory fit to low energy phenomenology.  It
gives an impressive account of the low energy gauge groups and matter
fields, but does less well in quantitative predictions.  Most
quantitative issues depend on where the world sits in moduli space.
More disturbingly, the theory gives a model independent prediction of
the unification scale in terms of the unified fine structure constant
and the Planck mass.  This prediction is off by a factor of $20$.
Furthermore, in order to get the right value of the unification scale,
one must choose the dimensionless string coupling to be of order at least
$10^7$.  The weak coupling theory does not give a consistent account of
phenomenology. 

Witten\ref\witten{E. Witten, ``Strong Coupling Expansion of
Calabi-Yau Compactification, hep-th/9602070, Nucl. Phys.
{\bf B. 471} (1996) 135.}
 has recently argued that the strong coupling limit.of heterotic
string theory, which is M theory on an interval\ref\horava{
P. Horava and E. Witten, ``Heterotic and Type I String Dynamics From
Eleven Dimensions'', {\it Nucl. Phys.} {\bf B460}, (1996),506, hep-th/9510209;
``Eleven-Dimensional Supergravity on a Manifold with
Boundary," hep-th/9603142.} gives a
better fit to the data.  In the M theory regime, the dimensionless
string coupling is interpreted as the size of the eleventh dimension,
measured in eleven dimensional Planck units, $\lambda \sim
(\Releven\Meleven)^{2/3}$.  The $E_8$ gauge groups live on the two ten
dimensional boundaries of the eleven dimensional manifold.  When one
group is broken by the standard imbedding of the spin connection of the
Calabi Yau manifold in the gauge connection, the boundaries act like
``capacitor plates'' for the eleven dimensional massless fields.  Witten
shows that this leads to a linear growth of the inverse $E_8$ coupling with
$\Releven$.  The coupling reaches infinity at a finite value of
$\Releven$ which is close to the value required to fit phenomenology.

Using formulas presented in \witten ,
one finds the following connections between the
$11$ dimensional Planck mass, $M_{11}$ (defined in terms of the
coefficient of the Einstein lagrangian in 11 dimensional supergravity, as
$M_{11}= \kappa_{11}^{-2/9}$), the $11$-dimensional radius,
$\Releven$, and the compactification radius, $R= V^{1/6}$, where
$V$ is the volume of the Calabi-Yau space on the boundary with unbroken
$E_6$ gauge group:
\eqn\rhosquared{\Releven^2 =  {\alpha_{GUT}^3 V \over 512 \pi^4 G_N^{2}
 },}
where $G_N$ is the four dimensional Newton's constant;
\eqn\meleven{M_{11}= R^{-1} \left (2 (4 \pi)^{- 2/3} \alpha_{GUT}
\right )^{-1/6}.}

Substituting $M_{GUT} = R^{-1} = 10^{16}$ GeV, $\alpha_{GUT} = {1\over
25}$ and the correct value for Newton's constant, one finds that
\eqn\melevenr{\Meleven R \sim 2}
\eqn\melevenreleven{\Meleven\Releven \sim 72}
In the paper\ref\coup{T.Banks, M.Dine, \lq\lq Couplings and Scales in
Strongly Coupled Heterotic String Theory\rq\rq , hep-th/9605136} on which
this talk is based, we chose the unification scale to be a factor of
three larger and thus had a smaller value of the size of the eleventh
dimension.   These values of the parameters are quite remarkable.  They
imply that the scale at which \lq\lq quantum gravity\rq\rq corrections
to field theoretic predictions become important is the unification
scale.  The four dimensional Planck scale is a low energy artifact, and
does not control the strength of these corrections.  Furthermore, at a
scale two orders of magnitude below the unification scale, physics
becomes $5$ dimensional.  Gravitational physics is certainly more
accessible to experiment in this regime than it is in weakly coupled string
theory. 

\newsec{General Consequences of the Strong Coupling Limit}

The paper \coup\ was devoted to exploring further general consequences of
the assumption that nature is described by strongly coupled heterotic
string theory.  The most striking of these is the emergence of a
solution to the strong CP problem.  String theory has a wide variety of
axion candidates.  However, in the weakly coupled regime there seem to
be mechanisms which give all of them potentials much larger than that
generated by QCD.  In the strong coupling regime this is no longer the
case.  To see this it is most convenient to exploit the large size of
the eleventh dimension and
pass to a four dimensional
effective theory via a five dimensional effective theory first worked
out by Antoniadis {\it et. al.}\ref\anton{I. Antoniadis,
S. Ferrara, T.R. Taylor, ```N=2
Heterotic Superstring and Its Dual
Theory in Five Dimensions,''
Nucl.Phys. {\bf B460} (1996) 489,
hep-th/9511108 }, following \ref\towns{M.Gunyadin, G.Sierra,
P.K.Townsend, Nucl. Phys. {\bf B242} (1984) 244;
Nucl. Phys. {\bf B253} (1985) 573.}.  

In the weakly coupled theory, there are a set of $h_{1,1}$ complex
moduli $T^a$ whose imaginary parts are potential axions.  However, the
associated Peccei-Quinn (PQ) symmetries are broken by world sheet
instantons.  The corresponding fields $Y^a$ 
in the strong coupling description, descend from five dimensional vector
superfields.  Their imaginary parts are the fifth components of vector
fields and the four dimensional PQ symmetries can be thought of as
arising from five dimensional \lq\lq gauge transformations \rq\rq with
gauge functions that do not vanish on the boundary of spacetime where
the standard model gauge fields live.  Thus all effects which break
these symmetries must involve this boundary in some way.  There are two
distinct mechanisms of PQ symmetry breaking.  The first involves effects
localized in the standard model boundary.  If QCD is the largest gauge
group on the boundary then this will be the dominant effect here.  We
can also have membrane instantons, membranes stretched between the two
boundaries and wrapped around two cycles in the Calabi Yau manifold.
These are the strong coupling remnants of world sheet instantons.  We
refer to them as remnants, because the large value of $\Releven$ gives
an exponential suppression of these effects.  They are totally
negligible compared to QCD.

The strongly coupled region of heterotic moduli space thus gives a
solution of the strong CP problem under fairly generic conditions.
If $h_{1,1} > 1$ it also predicts at least one extremely light axion
with a Compton wavelength of cosmological size.  
The approximate PQ symmetry which protects the mass of this axion also
ensures that it has only very tiny coherent couplings to matter\foot{T.B.
would like to thank K. Choi for a discussion of this point.}
Thus it is not ruled out by experiments on coherent long range forces.
Probes of spin dependent long range forces can now be seen as detecting
the topology of the internal Calabi Yau manifold.  

It is possible that such light axions exist even when $h_{1,1} = 1$.
The strongly coupled vacuum has {\it boundary moduli} which, in the
large volume limit can be thought of as moduli of the $E_8$ gauge bundle
on the standard model boundary.  Explicit orbifold calculations 
suggest that these have axionlike couplings to gauge fields.
Furthermore, since they live only on the standard model boundary
their potential should arise only from effects that involve this
boundary.  Furthermore, we should expect that their couplings to ordinary
matter are stronger by a factor of $\Releven\Meleven$ than the couplings
of the $Y^a$ fields described above.  This is a consequence of the fact
that the latter are defined as averages
over fields which live in the bulk of the eleven dimensional spacetime.
If such boundary axions exist, then the QCD axion will be 
a linear combination of them and of the Im $Y^a$, with the dominant
component being a boundary field.  The other linear combinations will be
superlight axions.

For cosmological reasons, the decay constant of the axion is of considerable
interest.  In the case that the axion is a bulk modulus, one would have
imagined that, like the gravitational field, its couplings were of order
$m_4^{-1}$ where $m_4 \sim 2 \times 10^{18}$ GeV, is the reduced Planck mass.
However, in weakly coupled heterotic string theory, Kim and Choi
\ref\choi{K. Choi, J.E.Kim, {\it Phys. Lett.} {\bf 165B}, (1985), 71.},
have shown that the decay constant of the model independent axion is 
smaller than the Planck mass by a factor $16\pi^2$.  We might anticipate
similar factors in the strong coupling region, so axion decay constants
as small as $10^{16}$ GeV do not seem unreasonable.  Similar factors in
the coupling of moduli whose mass comes from SUSY breaking interactions
might substantially alleviate the cosmological moduli problem.  In such
a context a scenario in which an axion with decay constant $\sim
10^{16}$ GeV is the dark matter in the
universe may be viable\ref\axion{T. Banks,
M.Dine, ``The Cosmology of Stringi Theoretic
Axions,'' hep-th/9608197.}.  Note further that if the
axion originates as a boundary modulus, it will have an even smaller
decay constant.   

The strongly coupled region of heterotic moduli space contains a large
number of different light scalars with a variety of masses and
couplings.  Early universe cosmology in this region is undoubtedly quite
complex and rich.  It is too early to say whether an acceptable
cosmological scenario emerges.

\newsec{Holomorphy as a Calculational Tool}

As mentioned above, Witten found a diverging gauge coupling at finite
$\Releven$ by performing a classical calculation in eleven dimensional
supergravity.  In \coup\ we pointed out that this result could be
reproduced by extrapolating existing weak coupling
calculations\ref\weak{K. Choi, J.E. Kim, {\it op. cit.}, L.Ibanez,
H.P.Nilles,  Phys. Lett. {\bf 180B} (1986), 354.}.  The gauge coupling function is a holomorphic
function of the conventional chiral superfields $S$ and $T^a$.  The
action is invariant under shifts of the imaginary part of these fields
by multiples of $2\pi$.  This means that up to corrections which are
exponentially small when the real parts of these fields are large (which
is the case in the strong coupling regime), the gauge coupling function
is a linear function of $S$ and $T^a$ and is therefore exactly computed
by the tree and one loop contributions in weak coupling perturbation
theory.  This computation gives 
\eqn\coupl{f_{6/8} = S \pm T^a \int {b_a \wedge F \wedge F \over 8 \pi^2 }}
for the $E_6$ and $E_8$ couplings. Here $b_a$ is a harmonic $(1,1)$ form
and $T^a$ the associated chiral superfield.  
 We see the blowup of the $E_8$
coupling when $S$ and $T^a$ are comparable.  Apart from this sign, the
calculation primarily determines the relation between the natural basis
of chiral superfields in the weak and strong coupling descriptions.
At strong coupling, the $E_8$ function is given simply by ${\cal S}$ a
chiral superfield whose real part is just the volume of the Calabi Yau
manifold on the $E_8$ boundary, measured in terms of eleven dimensional
Planck units.  The difference of the two couplings determines that $T^a$
is $Y^a$, the chiral superfield which descends from one of the $h_{1,1}$
vector multiplets of the five dimensional theory.  It is proportional to
$\Releven$ when written in eleven dimensional Planck units.

The holomorphic calculation breaks down as we enter the strong gauge coupling
regime ${\cal S} \sim 0$.  We can no longer neglect exponentials of
${\cal S}$ and the coupling may not really go to infinity.

Finally, we note that the calculation of PQ symmetry breaking referred
to above can also
be performed by analytic extrapolation of the weak coupling formula for
world sheet instanton effects.

\newsec{Supersymmetry Breaking and Modular Stabilization}

The most important phenomenological problem of string theory is to
understand the mechanism which stabilizes the moduli at fixed values.
This is true both because these values will determine the nature of low
energy physics and because unstabilized moduli are massless fields whose
properties are incompatible with a variety of astrophysical and
terrestrial observations.  The M theory regime sheds new light on the
stabilization problem, but does not solve it.

First of all, the region of large Calabi Yau volume and large $\Releven$
is a region of instability.  Once the moduli enter into this region they
tend to flow to infinity.  This was inevitable given the generality of
the analysis of \ref\dineseiberg{ M. Dine and N. Seiberg,
Phys. Lett. {\bf 162B}, 299 (1985),
and in {\it
Unified String Theories}, M. Green and D. Gross, Eds. (World Scientific,
1986)}. The phenomenological values of the
parameters and the analysis of \witten\ suggest some new possibilities
for stabilization at finite volume.
The linear size of the Calabi Yau manifold appears to be of order the
eleven dimensional Planck scale.  Thus, we should not trust a
semiclassical calculation of its Kahler potential.  Furthermore, the
five dimensional analysis of \anton\ shows that the Calabi Yau
volume is part of a five dimensional hypermultiplet.  Thus, even if
$\Releven $ is large, five dimensional symmetries will not fix its
Kahler potential.  These observations suggest that {\it
Kahler stabilization} of the volume modulus (and all others which come
from five dimensional
hypermultiplets ), which was advocated in \ref\banksdine{T. Banks and M. Dine,
``Coping with Strongly Coupled String Theory,''
hep-th/9406132, Phys. Rev. {\bf D50}
(1994) 7454.}, may be
operative here.  

In addition, Witten\witten\ finds that the
phenomenological values of the coupling are close
to the
point where the $E_8$ coupling seems to blow up.
In this region one can no longer argue
that the superpotential generated by gaugino condensation is a pure 
exponential, or indeed that the mechanism which generates the
superpotential can be understood on the basis of low energy physics.  
The superpotential is a complicated function of the moduli subject to
only mild symmetry constraints\foot{There may be some sophisticated
argument which in fact determines this complicated function.  However,
any such argument would depend on a knowledge of the singularities of
the function at finite values of its argument, and we are presently
unaware of any information about this.}
It is then reasonable to assume that all hypermultiplet moduli are
frozen at discrete values determined by nontrivial solutions of the
F-flatness conditions.  Note however that the nontrivial superpotential
cannot depend on the vector moduli $Y^a$.  Such a dependence is
forbidden by the approximate PQ symmetries, which are not broken by
strong coupling dynamics on the $E_8$ boundary.  

These assumptions lead immediately to an approximate no-scale model for SUSY
breaking at large $Y$.  Indeed, at large $Y$, five dimensional SUSY
fixes the Kahler potential of the $Y^a$, and the superpotential is
nonzero but does not depend on them.  The SUSY breaking scale is
$F \sim Y^{- 3/2} W$.  Presumably, since it contains no small coupling
parameters, the superpotential is of order the eleven dimensional Planck
scale.  Thus, for phenomenologically reasonable values of $Y$, the scale
of SUSY breaking is much too large.  

The vacuum energy predicted by this model is, barring unexplained
cancellations, of order $Y^{-2}$ times the square of the SUSY breaking
order parameter.   Thus, even if we retreat from the assumption of
strong coupling stabilization, and choose a weak coupling scenario in
which an exponentially small superpotential is stabilized at the proper 
scale of SUSY breaking by the Kahler potential, the vacuum energy is too
large to be compatible with observation.  Of course, we did not really
expect to solve the cosmological constant problem this easily.

Another generic prediction in this regime is that the leading order
squark mass matrix is of the same order as gaugino masses and is flavor
independent.  This follows from homogeneity of the leading order terms
in $Y$, for large $Y$.  It is unclear whether the corrections to this
result are small enough to account for the absence of flavor changing
neutral currents.   The real problem is that to leading order in $Y$
there is no stable minimum for the potential of $Y$ itself.  This is
another example of the general difficulty exposed in \dineseiberg .
Thus in order to find a satisfactory vacuum state we must contemplate 
competition between different orders in the $Y$ expansion.  This is
somewhat less plausible for the parameter values ($\Releven\Meleven \sim
200$) that we have chosen in this lecture than for those used in \coup .
But even if we find a stable vacuum at a large value of $Y$, its very
existence leads us to doubt the reliability of the expansion.  Thus, we
can have little confidence in our prediction of squark degeneracy.

\newsec{Conclusions}

Strongly coupled heterotic string theory is a better fit 
to the parameters of nature than the weakly coupled version.
If correct, it implies potentially dramatic gravitational physics at
energy scales well below the Planck scale.  Conventional approaches
to grand
unification and to inflation may have difficulty surviving in this
environment.  The nominal scale at which quantum gravitational
corrections become important is of order both the unification scale and
the scale of energy density in the simplest inflationary models.
Moreover, the strongly coupled heterotic theory implies that the radius
of the fifth dimension is one or two orders of magnitude larger than
the inverse unification mass.  

In the strongly coupled regime, string theory provides a solution of the
strong CP problem and may also predict one or more superlight axion
fields.  The value of the axion decay constant and the cosmological
implications of the theory are under investigation.  

There is also an interesting new slant on the problems of modular
stabilization and supersymmetry breaking.  Strong coupling physics on
the \lq\lq hidden boundary\rq\rq can lead to stabilization of many of
the moduli.  However, an approximate five dimensional supersymmetry
then implies too large a scale of $N=1$ SUSY breaking in the four
dimensional effective theory, via the no scale mechanism.  
This can be avoided by invoking Kahler stabilization of the moduli at a
point where the hidden sector gauge theory is weakly coupled.  
One then obtains a no scale model of SUSY breaking with approximately
degenerate squarks.  However, in the same approximation it is impossible
to stabilize the moduli fields which are remnants of five dimensional
vector multiplets.  This observation puts the entire scheme of SUSY
breaking under a cloud of suspicion.

Although the strong coupling theory is by no means ruled out by these
considerations, it clearly has many problems.  One is led to ask whether
its attractive features might be found in a wider class of string vacua.
The key feature which enabled Witten to obtain the required discrepancy
between the four dimensional Planck scale and the unification scale was
that the gauge sector of the low energy theory lives on a lower
dimensional submanifold of spacetime, while gravity propagates in bulk.
Recent work on string duality has shown that it is quite typical to have
gauge fields living on low dimensional manifolds called D-branes.  Thus,
there may be many classical string vacua which will naturally explain
the ratio of the Planck and unification scales.  Notice that it will
always be the case in such a vacuum that quantum gravity effects become
important below the Planck scale.   It is also fairly common to find
distortions of bulk gauge symmetries on D branes.  In an effective
four dimensional theory, these will show up as PQ symmetries broken by
nonperturbative physics attached to the brane.  Thus our observation
that the strong CP problem is solved in the strongly coupled heterotic
vacuum may generalize as well.  

We are thus motivated to search for general realizations of the field
content of the standard model in D-brane physics.  Given a list of such
constructions one could then try to embed them into full string
compactifications in which all tadpoles/anomalies are cancelled, and
search for examples with sensible dynamics for stabilizing the moduli
and breaking SUSY.

\listrefs
\bye
\end

\end